\documentclass[conference]{IEEEtran}

\IEEEoverridecommandlockouts

\usepackage{cite}

\usepackage[english]{babel} 

\ifCLASSINFOpdf
   \usepackage[pdftex]{graphicx}

\else

   \usepackage[dvips]{graphicx}

   \DeclareGraphicsExtensions{.eps}
\fi

\usepackage{grffile}
\usepackage{epstopdf}

\usepackage[cmex10]{amsmath}

\usepackage{color}

\begin{document}

\title{Average Power Limitations in Sliding Window Contention Resolution Diversity Slotted Aloha}

\author{\IEEEauthorblockN{Alessio Meloni\thanks{A. Meloni gratefully acknowledges Sardinia Regional Government for the financial support of his PhD scholarship (P.O.R. Sardegna F.S.E. 2007-2013 - Axis IV Human Resources, Objective l.3, Line of Activity l.3.1.).} and Maurizio Murroni\thanks{\copyright 2013 IEEE. The IEEE copyright notice applies. DOI: 10.1109/IWCMC.2013.6583526}}
\IEEEauthorblockA{DIEE - Dept. of Electrical and Electronic Engineering\\
University of Cagliari\\
Piazza D'Armi, 09123 Cagliari, Italy\\
Email: \{alessio.meloni\}\{murroni\}@diee.unica.it}
}


\maketitle
\begin{abstract}
Recently a new Random Access technique based on Aloha and using Interference Cancellation (IC) named Sliding Window Contention Resolution Diversity Slotted Aloha (SW-CRDSA) has been introduced. Differently from classic CRDSA that operates grouping slots in frames, this technique operates in an unframed manner yielding to better throughput results and smaller average packet delay with respect to frame-based CRDSA. However as classic CRDSA also SW-CRDSA relies on multiple transmission of the same packet. While this can be acceptable in systems where the only limit resides in the peak transmission power, it could represent a problem when constraints on the average power (e.g. at the transponder of a satellite system) are present. In this paper, a comparison in terms of normalized efficiency is carried out between Slotted Aloha and the two CRDSA techniques.\\ 
\end{abstract}

\begin{IEEEkeywords}
Random Access, Slotted Aloha, Interference Cancellation, Contention Resolution Diversity Slotted Aloha, Sliding Window, Power Limitations, Satellite Communications.
\end{IEEEkeywords}

\IEEEpeerreviewmaketitle
\section{Introduction}

Random Access techniques such as Slotted Aloha (SA) \cite{SWCRDSA:RobertsALOHA} \cite{SWCRDSA:AbramsonALOHA} and Diversity Slotted Aloha (DSA) \cite{SWCRDSA:DiversityALOHA} have been largely used especially in satellite communications both for initial terminal login and when small amounts of data need to be sent. Their almost 40 years long success resides, among the others, in their capability to work nicely in peculiar conditions such as long propagation delay and directional transmissions that do not allow transmitting terminals to have an immediate feedback either about the state of the channel in terms of occupancy or about the outcome of their transmission. This is true especially in the case of Geostationary Earth Orbit (GEO) satellites with bent pipe repeaters, for which the Round Trip Time (RTT) between terminals and the gateway is approximately $500\ ms$.
However, the absence of coordination among users introduces the possibility of collision among bursts sent from different users and the subsequent loss of the transmitted content. For this reason, ALOHA-based techniques have been generally used when the expected load on the channel is small enough to ensure a sufficiently low packet loss probability.  

Recently, Aloha-based techniques have gained increasingly new attention. This is due to the introduction of the concept of Interference Cancellation (IC) as a mean to exploit the diversity advantages brought by DSA. In particular this new technique, called Contention Resolution Diversity Slotted ALOHA (CRDSA), was first introduced in \cite{SWCRDSA:CRDSA1} and allows to restore the content of colliding packets based on the fact that if two identical copies of the same packet are sent and each one contains a pointer to the position of the other one, the interference contribution due to one copy can be removed in case the other copy is correctly decoded. Subsequently, the same concept has been extended to more than two copies per packet \cite{SWCRDSA:CRDSA2} \cite{SWCRDSA:CRDSA3} and to the case of variable burst degree, known as Irregular Repetition Slotted Aloha (IRSA) \cite{SWCRDSA:IRSA1}, in which for each packet a certain number of copies are sent according to a given probability distribution.
Let us define $G$ as the normalized MAC channel load, i.e. the average number of different packet contents sent per frame normalized over the frame size and $T=G (1-PLR)$ as the throughput, with PLR standing for Packet Loss Ratio. While for SA the maximum throughput value is $T\simeq 0.37 [pkt/slot]$ (obtained for $G=1$) and DSA ensures smaller packet loss probability up to moderate loads, CRDSA and its evolutions can reach throughput values even close to $1 [pkt/slot]$. 
As a matter of fact, original CRDSA with 2 copies per packet can get to $T\simeq 0.55 [pkt/slot]$, for CRDSA with more than 2 copies $T\simeq 0.7 [pkt/slot]$ and CRDSA with Variable Burst Degree can achieve $T\simeq 0.938 [pkt/slot]$ if the maximum allowed packet repetition is equal to 8.\footnote{This values are upper bounds that can be reached only with an asymptotic setting as claimed in \cite{SWCRDSA:IRSA1}. Peak values obtained for a realistic frame size will be shown in Section~\ref{OW}.}

However, in CRDSA all the replicas of the same packet are placed within the $N_S$ slots of a frame. This implies that each user has to wait the beginning of a new frame to start sending its content. Therefore, a new and undesirable component of delay is introduced with respect to SA in which a packet is typically sent in the next slot as soon as the content is ready for transmission.
Also the throughput performance is limited by frames since packets sent in the same frame share the same set of eligible slots to place their copies and this increases, from a probabilistic point of view, the occurrence of unsolvable collisions. For this reason, in \cite{SW} a new technique exploiting the advantages of CRDSA in an unframed manner has been introduced. This technique (named Sliding Window - Contention Resolution Diversity Slotted Aloha) further boosts the throughput performance up to 13\% with respect to CRDSA and at the same time reduces the average packet delay at destination.
Nevertheless, also Sliding Window - CRDSA (in short SW-CRDSA) similarly to classic CRDSA (from now on indicated as FB-CRDSA standing for Frame-Based CRDSA) relies on the transmission of multiple copies of the same packet per attempt. While this can be acceptable in systems where the only limit resides in the peak transmission power, it could represent an issue when a constraint on the average power is present as for example at the satellite transponder relaying terminals data to a remote gateway. For this reason in the following paper, similarly to what has been done in \cite{SWCRDSA:IRSA1} for FB-CRDSA, an analysis and comparison in terms of normalized efficiency is carried out for SW-CRDSA and obtained results are compared with the normalized efficiency in case of SA and FB-CRDSA.

The paper is organized as follows. In Section II an overview of the considered access scheme is presented. In Section III the reasoning behind the computation of the normalized efficiency as well as the formulas used for the analysis are given. Section IV illustrates simulation results. Section V deals with remarks on the energy efficiency aspects when retransmissions are taken into account. Section VI concludes the paper.

\section{Access scheme overview}\label{OW}

\subsection{Transmitter side}

\begin{figure}[h!]
\includegraphics [width=\columnwidth] {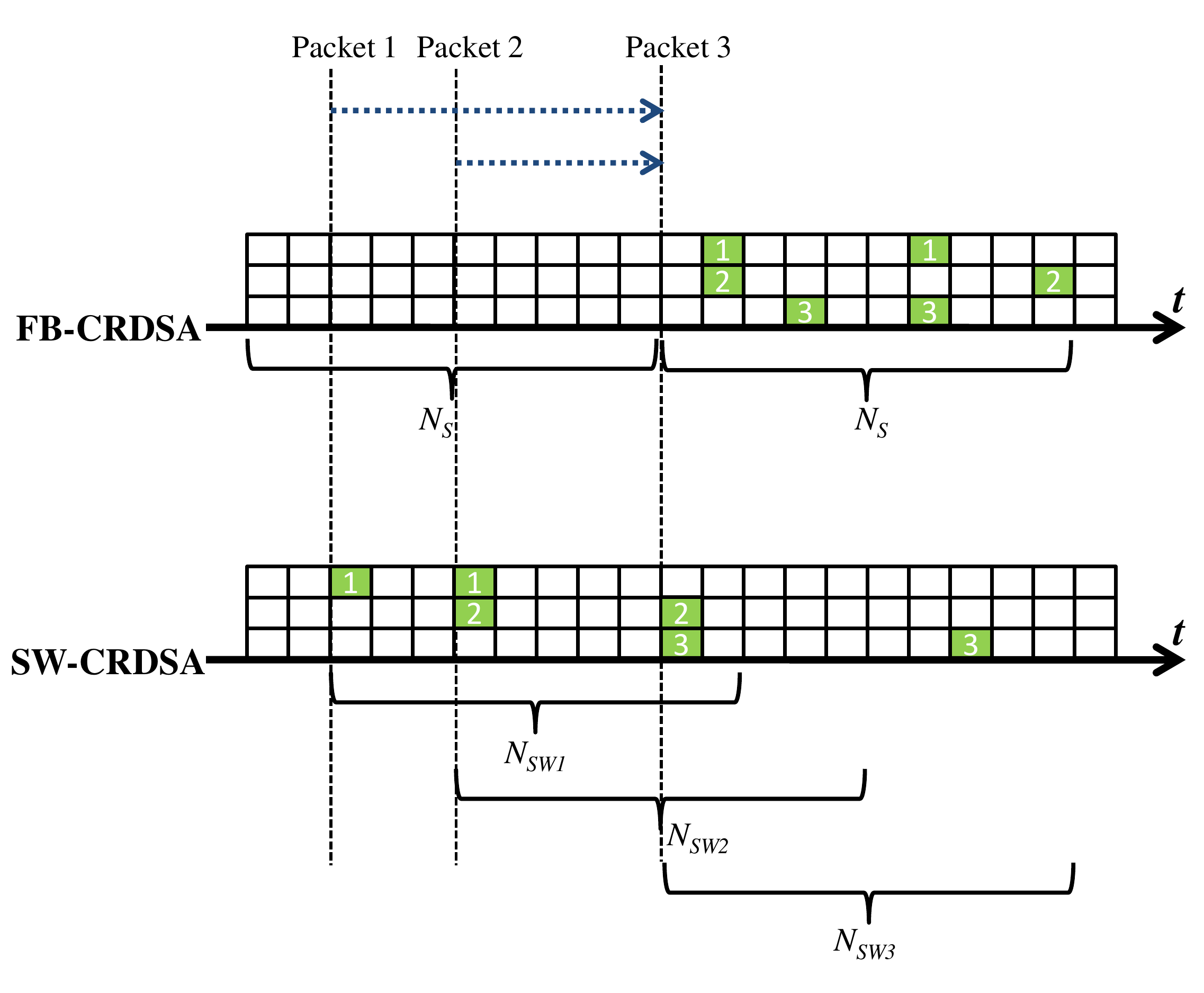}
\caption{Example of access to the channel for FB-CRDSA and SW-CRDSA.}
\label{AxSchemes}
\end{figure}

Consider a scenario in which a certain number of terminals communicate to a remote gateway via satellite using SC-TDMA and have no immediate knowledge either about the outcome of their transmission or about the status of the other terminals (transmitting or not). Figure~\ref{AxSchemes} shows an example of how FB-CRDSA and SW-CRDSA differently behave when a certain number of new packets are ready for transmission. The instant in which packets are ready for transmission is indicated as a vertical line. In this example the slot time has been assumed as time unit so that packet arrivals always occur at slot starts. Although this approximates the real case, this approximation does not substantially conditions our results since the slot time $T_S$ is much smaller than the frame interval $N_S\cdot T_S$ and the only change regards delay results. 

As done in \cite{SWCRDSA:IRSA1}, let us define the burst degree distribution (i.e. the probability of having a certain number of copies per packet) through the following polynomial representation\\

\begin{equation}
\Lambda(x) = \sum_l \Lambda_l x^l
\end{equation}\\ 

where $\Lambda_l$ is the probability that a given packet will have burst degree $l$. 
While in FB-CRDSA each packet copy is placed within the $N_S$ slots of the next starting frame (with equal probability for a slot to contain a copy of a given packet), in SW-CRDSA the first copy is sent immediately in the next starting slot while the other $l-1$ copies for the same packet are placed in the next $N_{sw}-1$ slots with equally distributed probability. The set of $N_{sw}$ slots (called Sliding Window\footnote{The name Sliding Window comes from the fact that depending on the moment packet copies have begun to be sent, the set of slots to be considered is gradually sliding in time.}) represents the number of successive slots, comprehensive of the one with the first packet copy, in which a certain user places all the replicas of a given packet and in this sense it can be considered as counterpart of the FB-CRDSA frame.
Therefore, in FB-CRDSA packets have an additional delay component that varies from $0$ to $N_S\cdot T_S$ depending on the moment they are ready for transmission. In Figure~\ref{AxSchemes}, this waiting interval is indicated with a dotted arrow that lasts from the time the packet was ready for transmission until the moment in which the related terminal is actually allowed to access the channel to send that packet.

\subsection{Receiver side}

At the receiver, assuming perfect channel estimation and interference cancellation, one of three possible situations can occur for each slot:

\begin{itemize}
\item{no burst copies are received;}
\item{only 1 burst is received, thus the packet can be correctly decoded and the content of the other copies belonging to the same packet can be removed from the other slots in order to unlock other packet contents;}
\item{more than 1 burst is received in the same slot, thus interference occurred and all the contents belonging to different terminals that are present in that slot cannot be decoded at the receiver side.}
\end{itemize}

\begin{figure}[t!]
\includegraphics [width=\columnwidth] {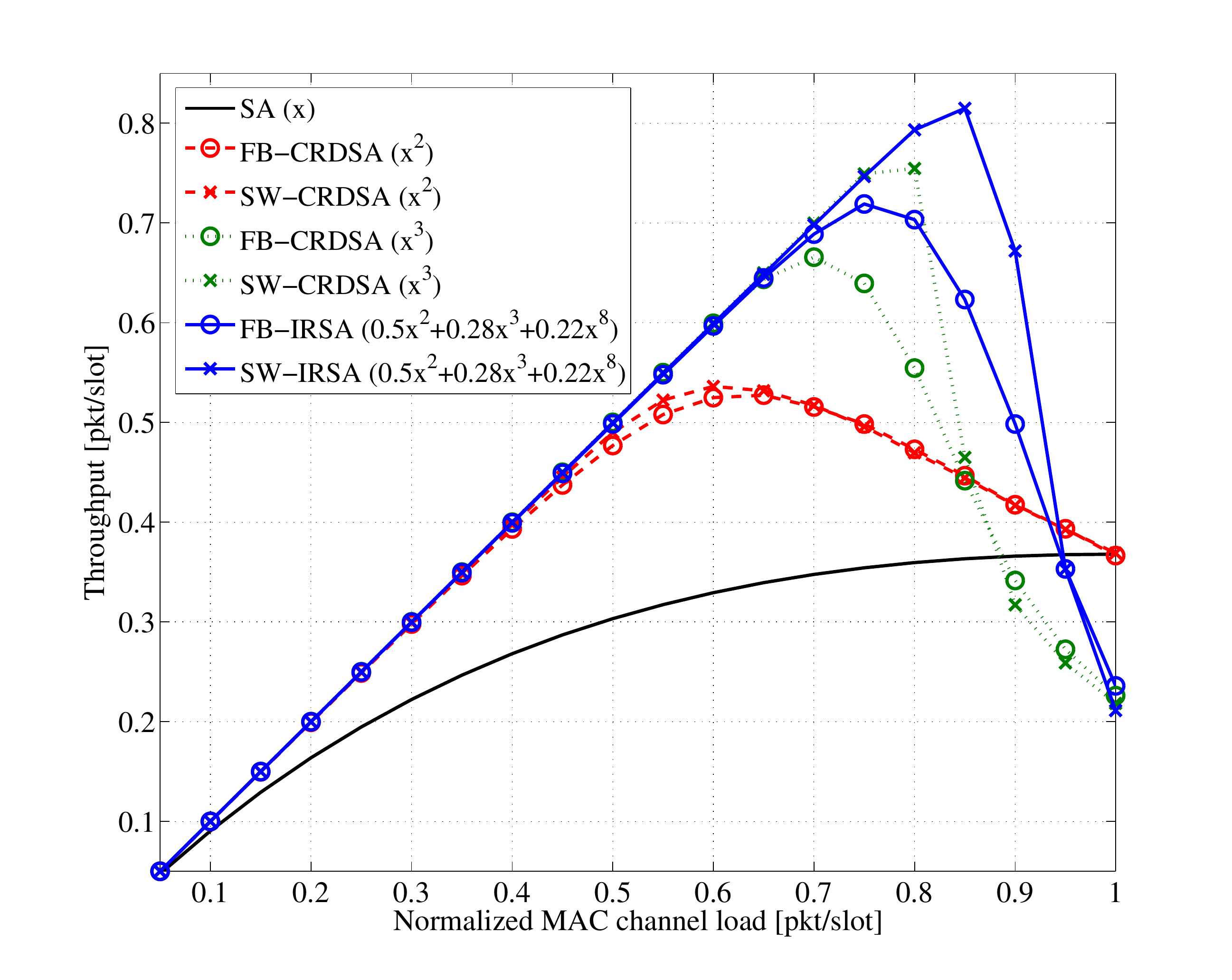}
\caption{Throughput for FB-CRDSA and SW-CRDSA with $I_{max}=50$ and $N_{sw}=N_s=200\ slots$.}
\label{TOTHRP}
\end{figure}

Based on the rules listed above, an iterative IC process is started at the receiver so that at each iteration the copies belonging to successfully decoded packets are removed from the other slots. Doing that, previously undecodable packets gain the possibility to be decoded if all the other bursts colliding in the same slot have been correctly decoded. Therefore the IC process allows recovery of further packets than those having at least one copy received without interference as in DSA. This process goes on until all possible packets have been decoded or until the maximum number of iterations $I_{max}$ is reached. Figure~\ref{TOTHRP} shows throughput results for various burst degree distributions both in the case of FB-CRDSA and in the case of SW-CRDSA assuming same maximum number of iterations for the IC process and same size for the sliding window and the frame ($N_{sw}=N_f$). Displayed results assume Poisson Arrivals for packets transmitted and $G$ represents the mean of the corresponding Poisson distribution\footnote{The assumption of Poisson Arrivals is necessary in order to obtain comparable results for FB-CRDSA and SW-CRDSA as thoroughly explained in \cite{SW}.}. 

As claimed in the introduction and demonstrated in \cite{SW}, the throughput for SW-CRDSA is generally higher than the one for FB-CRDSA if the same load for the two is considered. This can be explained with the fact that each packet's sliding window is different from the others unless more than one packet was ready for transmission within the same slot interval. Therefore the probability of unsolvable collisions depends on the moment in which packets are ready for transmission (differently from FB-CRDSA) and in particular, if the number of simultaneous packet transmission starts is sufficiently low, the probability of unsolvable collisions is smaller with respect to FB-CRDSA.
This comes at the cost of the need to keep a bigger number of slots in memory at the receiver for the iterative IC process. In fact, in FB-CRDSA all the solvable collisions are bounded within one frame. Therefore, after the decoding process for a given frame is finished and decoded packets have been passed to the upper layers, memorized slots are not anymore useful and can thus been removed if needed. In SW-CRDSA instead, also bursts received more than $N_{sw}$ slots before have the possibility to still be correctly decoded due to the fact that interaction among packet copies can not be a-priori bounded within a certain number of slots. For this reason, the need to keep in memory a number of slots greater than $N_{sw}$ arises. While this fact would theoretically require an infinite memory, it has been demonstrated in \cite{SW} that keeping in memory at the receiver $(5\cdot N_{sw})\ slots$ it is sufficient to avoid almost any loss of potentially decodable packets.

\begin{figure}[t!]
\includegraphics [width=\columnwidth] {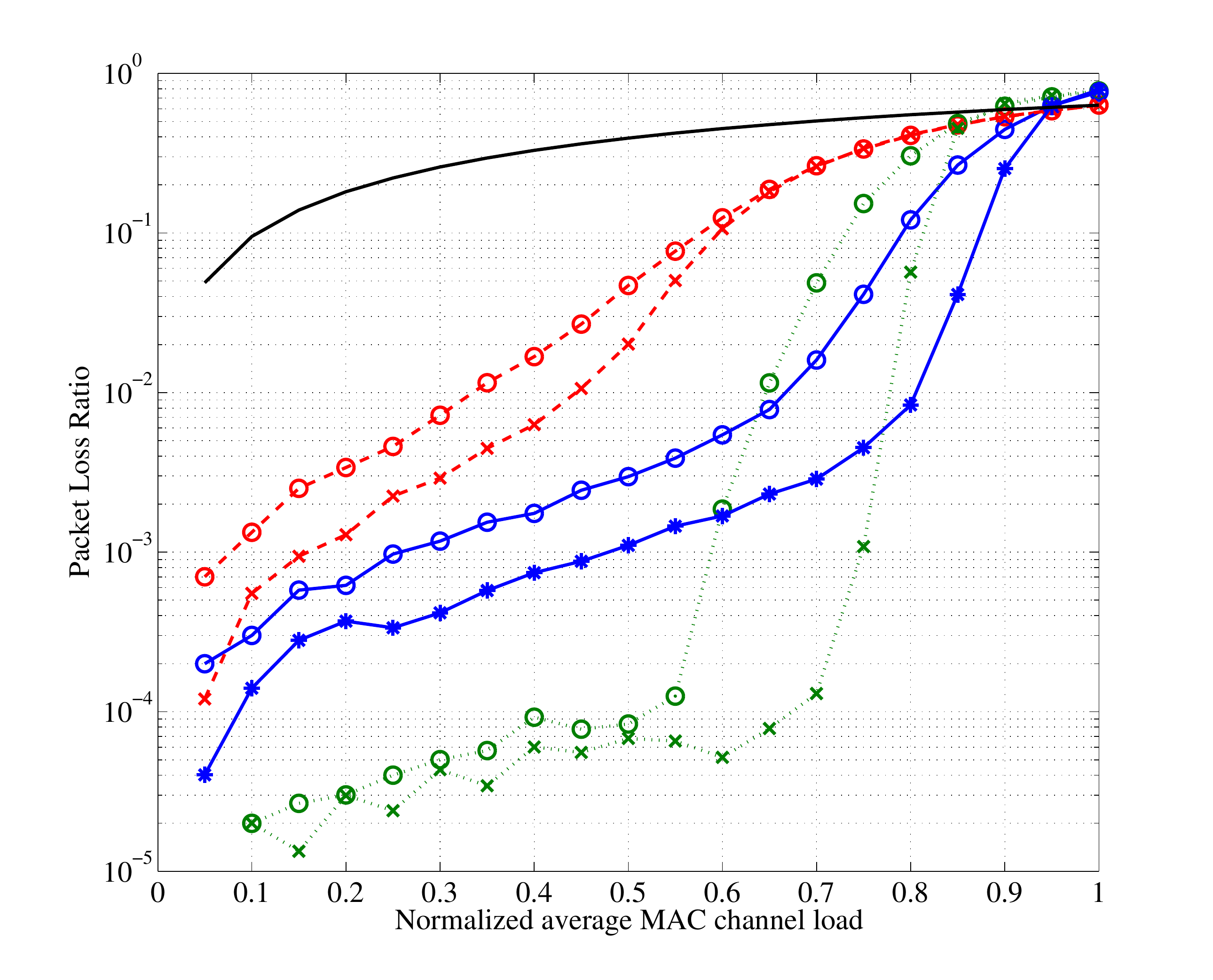}
\caption{Packet Loss Ratio for FB-CRDSA and SW-CRDSA with $I_{max}=50$ and $N_{sw}=N_s=200\ slots$.}
\label{PLR_TOTAL}
\end{figure}

Finally a comparison in terms of Packet Loss Ratio is shown in Figure~\ref{PLR_TOTAL}. As we can see, also in this case SW-CRDSA outperforms FB-CRDSA thus justifying the advantage of its use also for small loads. To be noticed also that while for the peak throughput the best results are obtained for Variable Burst Repetition, in this case the communication better benefits from the use of a regular number of replicas.

\section{Normalized efficiency}

The comparison presented in the previous section assumes the same peak transmitting power for all schemes. As already pointed out in \cite{SWCRDSA:AbramsonALOHA} and \cite{SWCRDSA:IRSA1}, this assumption is correct for many applications in which the only limit resides on the peak power available and the main interest regards the effect of the interference due to multiple access. However there are cases in which a limit on the average power is present. For example, in satellite systems the average power available at the transponder represents a fundamental limitation for transmission in the downlink path (i.e. from the satellite to the earth receiver). For this reason, it is of interest to analyze the throughput of this new medium access scheme assuming the same average signal power received at the earth station.

To do so, we consider the normalized efficiency $\eta$, defined similarly to \cite{SWCRDSA:AbramsonALOHA} as the ratio of the capacity $C_i$ (with $i$ indicating the considered Random Access Scheme) to the Gaussian capacity $C_{ref}$ (i.e. the capacity of the satellite channel under the assumption that the transponder transmits continuously):

\begin{equation}
\eta=\frac{C_i}{C_{ref}}
\end{equation}

The Gaussian channel capacity $C_{ref}$ is expressed as

\begin{equation}
C_{ref}=W\cdot log\Bigg(1+\frac{P}{N}\Bigg)
\end{equation}

where $W$ is the channel bandwidth, $P$ is the average aggregate signal power at the receiver and $N$ is the noise power. Moreover, from \cite{SWCRDSA:IRSA1} the capacity of the considered RA scheme can be evaluated as

\begin{equation}
C_{i}=W\cdot T_i(G)\cdot log\Bigg(1+\frac{P}{N\cdot D_i}\Bigg)
\end{equation}

where $G$ is the normalized MAC channel load, $T_i(G)$ the related throughput and $D_i$ is the ratio between the average transmitted power and the power used for the transmission of a packet copy. Therefore in SA $D_{SA}=G$, in CRDSA with a regular number $l$ of replicas $D_{CRDSA}=(l\cdot G)$ and in the more general case of irregular repetitions $D_{IRSA}=(\Lambda'(1)\cdot G)$ where $\Lambda'(1)$ is the average burst degree as defined in \cite{SWCRDSA:IRSA1}.

\section{Simulation Results}

\begin{figure}[t!]
\centering
\includegraphics [width=\columnwidth] {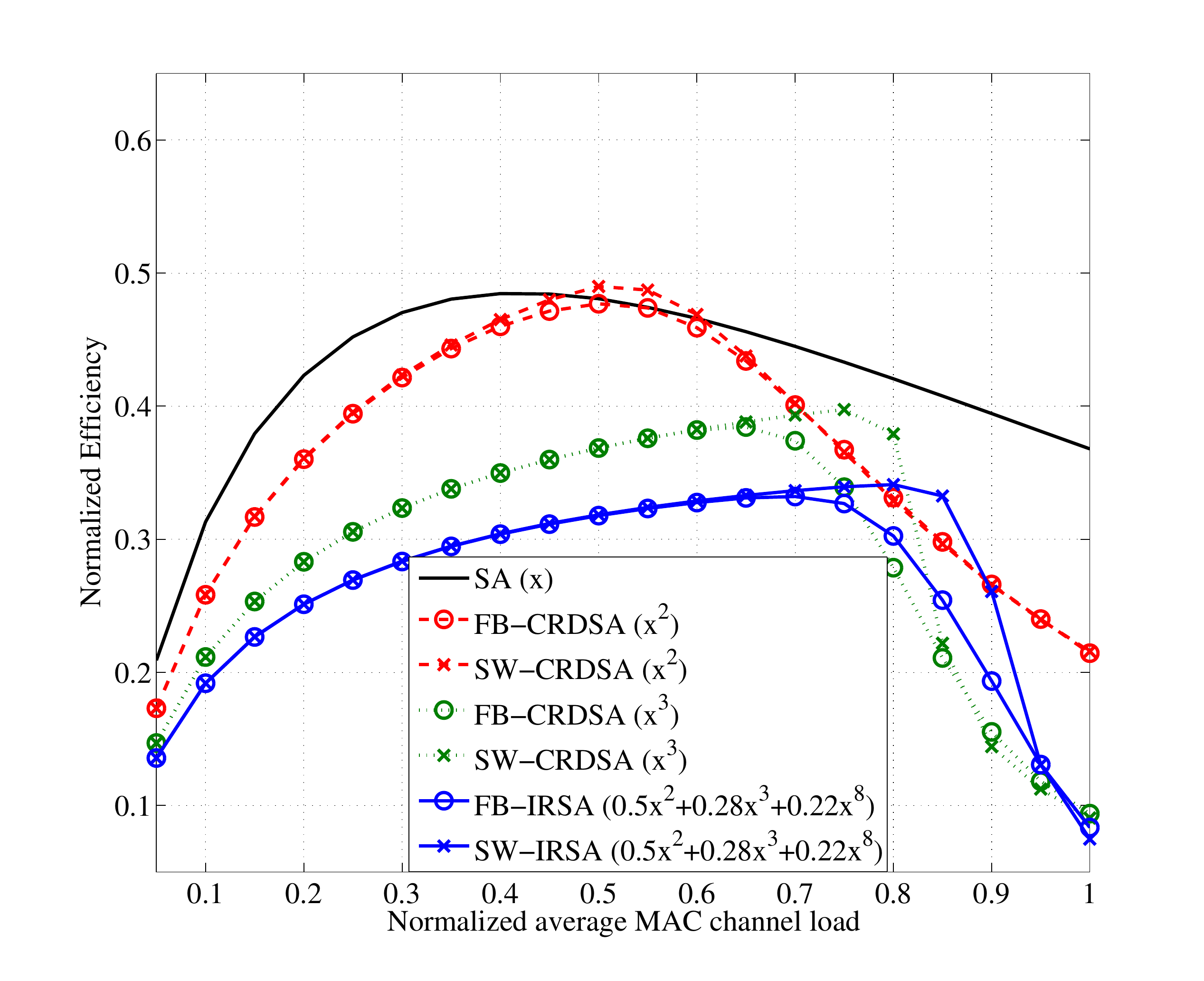}
\caption{Normalized Efficiency for SA and various Frame Based and Sliding Window packet replicas distributions with  $N_f=N_{sw}=200$ slots, $I_{max}=50$ and $SNR=0\ dB$.}
\label{nef0}
\end{figure}

\begin{figure}[h!]
\centering
\includegraphics [width=\columnwidth] {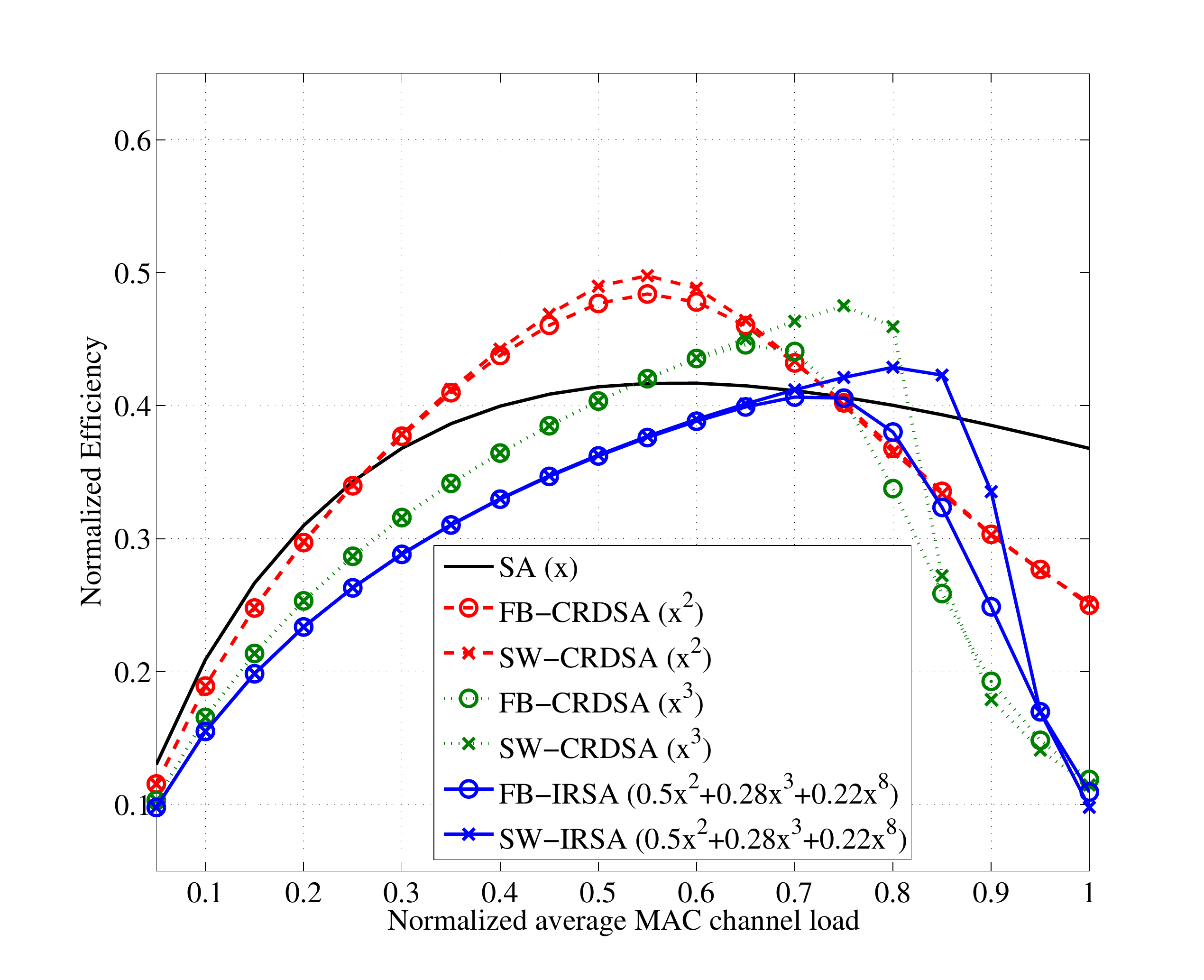}
\caption{Normalized Efficiency for SA and various Frame Based and Sliding Window packet replicas distributions with  $N_f=N_{sw}=200$ slots, $I_{max}=50$ and $SNR=6\ dB$.}
\label{nef6}
\end{figure}

\begin{figure}[h!]
\centering
\includegraphics [width=\columnwidth] {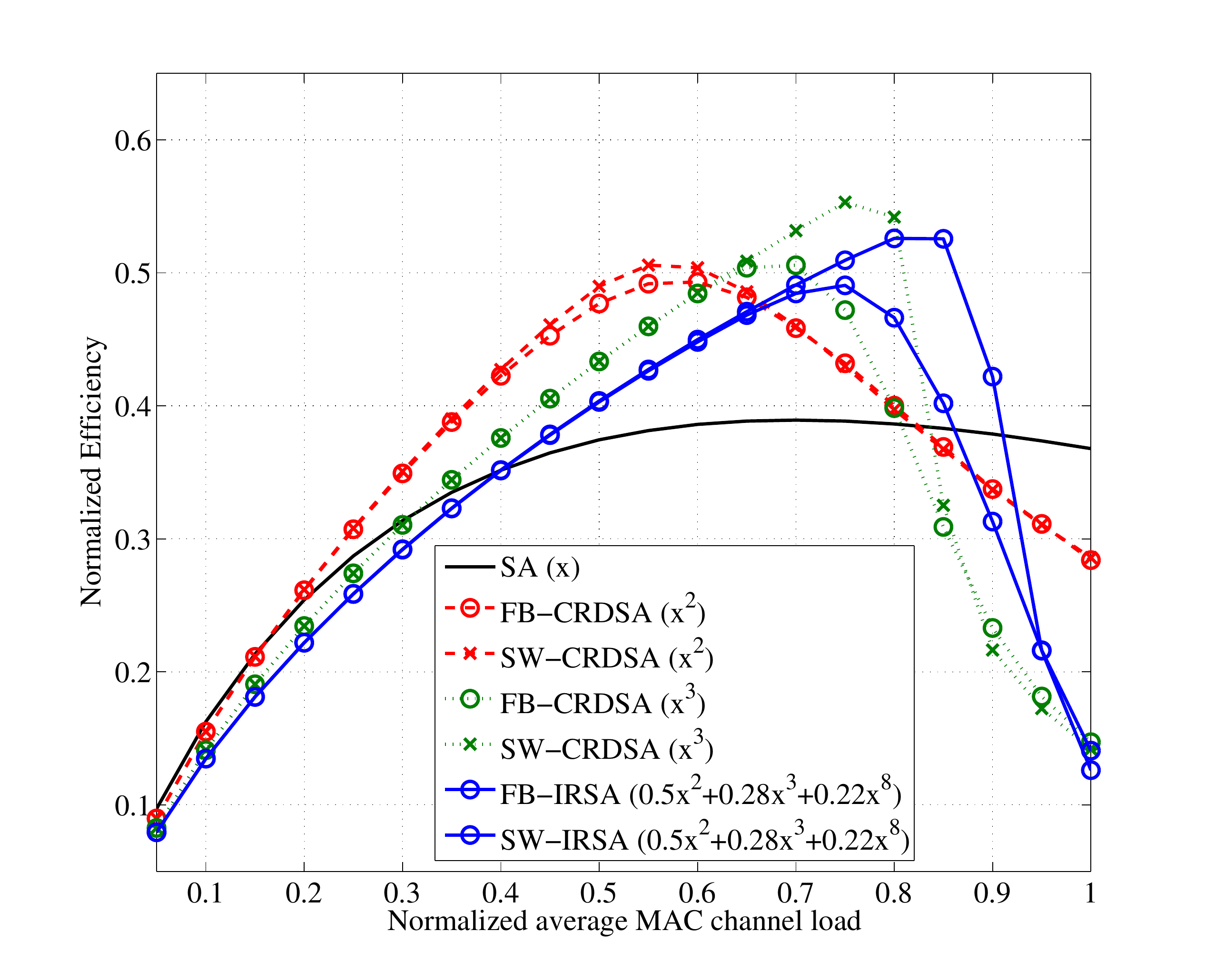}
\caption{Normalized Efficiency for SA and various Frame Based and Sliding Window packet replicas distributions with  $N_f=N_{sw}=200$ slots, $I_{max}=50$ and $SNR=12\ dB$.}
\label{nef12}
\end{figure}

\begin{figure}[h!]
\centering
\includegraphics [width=\columnwidth] {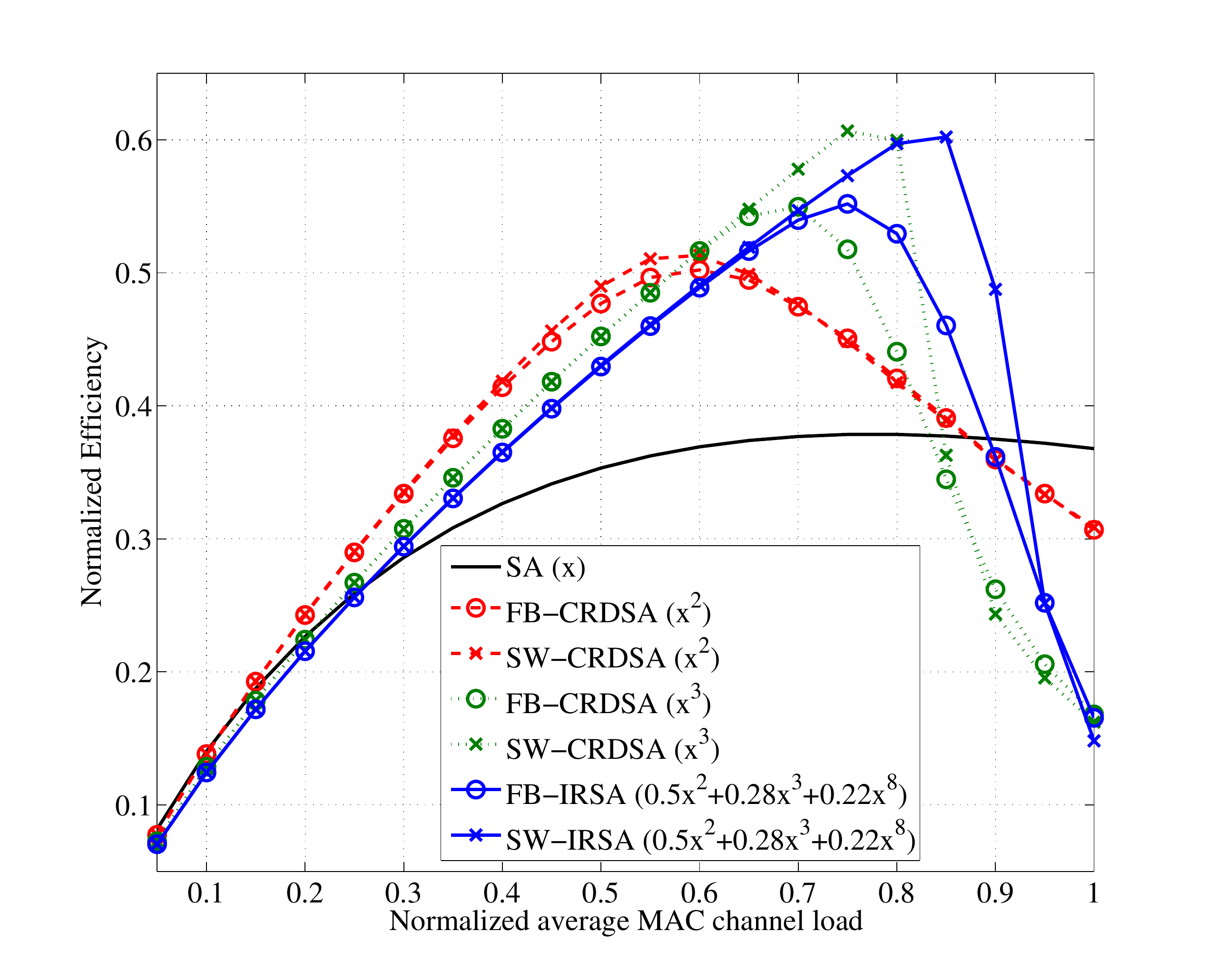}
\caption{Normalized Efficiency for SA and various Frame Based and Sliding Window packet replicas distributions with  $N_f=N_{sw}=200$ slots, $I_{max}=50$ and $SNR=18\ dB$.}
\label{nef18}
\end{figure}

Based on the access scheme overview given in Section II and on the definition of normalized efficiency given in Section III, in this section simulation results in terms of normalized efficiency depending on the normalized MAC channel load (i.e. logical channel load regardless of the actual physical number of bursts per packet content) are shown for various $SNR$ values and under the constraint of equal average power at the receiver. The following simulations have been obtained through implementation in a numerical computing environment, assuming for each point of the resulting curve that the total arrivals of packets to be transmitted are Poisson distributed with aggregate channel load value per each point equal to the mean value of the corresponding Poisson distribution. As already outlined in Section II, a typical scenario where these simulations could be applied is the case of a certain number of terminals that send bursty and infrequent data to a remote gateway via satellite using SC-TDMA.
Figures~\ref{nef0}~-~\ref{nef18} show that the obtained results are highly dependent on the utilized burst degree distribution as well as on the SNR. While from a general point of view we can immediately state the convenience in using SW-CRDSA instead of FB-CRDSA, a more in-depth analysis on the best burst degree distribution needs a discussion of the presented figures.  

Figure~\ref{nef0} shows that for $SNR=0\ dB$, FB-CRDSA with 2 replicas gets worse results than SA and only equals it in terms of normalized efficiency for $G=0.55$. SW-CRDSA with 2 replicas instead, outperforms SA both in terms of normalized efficiency in the range between $G=[0.45,0.6]$ and in terms of normalized efficiency peak. The nice thing about SW-CRDSA outperforming SA precisely in this range comes from the fact that this is the region around the throughput peak, i.e. the area in which we want our communication system using CRDSA as Random Access method to operate from a throughput maximization perspective. The use of other burst degree distributions than $\Lambda(x)=x^2$ yields to bad results over the entire range of load values, compared to Slotted Aloha.

For $SNR=6\ dB$, the convenience of using CRDSA($x^2$) becomes more and more evident while also the choice of a greater number of replicas is found to be a better choice with respect to SA if the operating point is around the throughput peak. However, at $SNR=6\ dB$ the use of more than 2 replicas per packet still does not appear to be the best choice with respect to CRDSA($x^2$).

Finally for $SNR=12\ dB$ using SW-CRDSA($x^3$) becomes the best choice while from $SNR=18\ dB$ SW-IRSA with maximum burst degree equal to 8 begins to outperform the normalized efficiency of CRDSA with regular burst distribution.

\section{Remarks on normalized efficiency in the case of retransmissions}

The analysis carried out so far assumes energy fairness in an open loop scenario, i.e. in the case in which only a single transmission attempt per packet content takes place. However, there are a number of other scenarios in which retransmission of failing packets is required. The presented analysis can be extended to those systems under the constraint that the ongoing communication is stable in terms of overall channel load generated by terminals. In the followings a generalization of the definition of stability given in \cite{CRDSA_stab1} is presented and the rationale for stating the validity of the analysis in case of retransmissions is given. 

Consider a certain population of users $M$ that participate in a communication scenario using one of the techniques described so far. Users can be either in non-backlogged or in backlogged state. Assuming that users can handle no more than one packet at the time, in non-backlogged state users are idle because they do not have any packet to transmit or they are awake and waiting for starting a new transmission according to a certain transmission policy $p^{tx}$; in backlogged state, users want to retransmit a packet that has not been correctly received and attempt a retransmission according to a given policy $p^{retx}$. Therefore, the total load present in the channel depends on the load due to new transmissions $G_{tx}$ (determined by the number of non-backlogged users and the transmission policy $p^{tx}$) and on the load due to retransmissions $G_{retx}$ (determined by the number of backlogged users and the retransmission policy $p^{retx}$). The sum of these two quantities $G=G_{tx}+G_{retx}$ constitutes the normalized MAC channel load $G$.
The requirement needed in order to consider the open loop analysis still valid is that the expected channel load remains the same over time. Considering $p^{tx}$ and $p^{retx}$ to be stationary policies, the expected channel load will be the same over time if the expected number of backlogged users remains the same\footnote{By reflection this means that also the number of non-backlogged users remains equal since the number of non-backlogged users is by definition the total population $M$ minus the number of backlogged users $N_B$.}. This corresponds to the requirement that for a certain number of users switching to backlogged state, an equal number of users in backlogged state switches back to non-backlogged state so that  $G_{tx}(N_B^*)=T(N_B^*)$ and the communication can be considered to be in a point of equilibrium for $N_B^*$ backlogged users.
In particular, considering an arbitrarily small positive quantity $\epsilon$, the equilibrium point is of stable equilibrium if the neighborhoods of the equilibrium point are such that 
\begin{equation}
G_{tx}(N_B^*-\epsilon)>T(N_B^*-\epsilon)
\end{equation}
and
\begin{equation}
G_{tx}(N_B^*+\epsilon)<T(N_B^*+\epsilon)
\end{equation}
i.e. the point of equilibrium acts as a sink. On the other hand, if
\begin{equation}
G_{tx}(N_B^*-\epsilon)<T(N_B^*-\epsilon)
\end{equation}
and
\begin{equation}
G_{tx}(N_B^*+\epsilon)>T(N_B^*+\epsilon)
\end{equation}
the point is of unstable equilibrium since it acts as a source.
If the former case is verified and the point of equilibrium is also the only one, it can be claimed that the point of stability is global and the communication will always have the same value of expected load.

\begin{figure}[h!]
\centering
\includegraphics [width=0.9 \columnwidth] {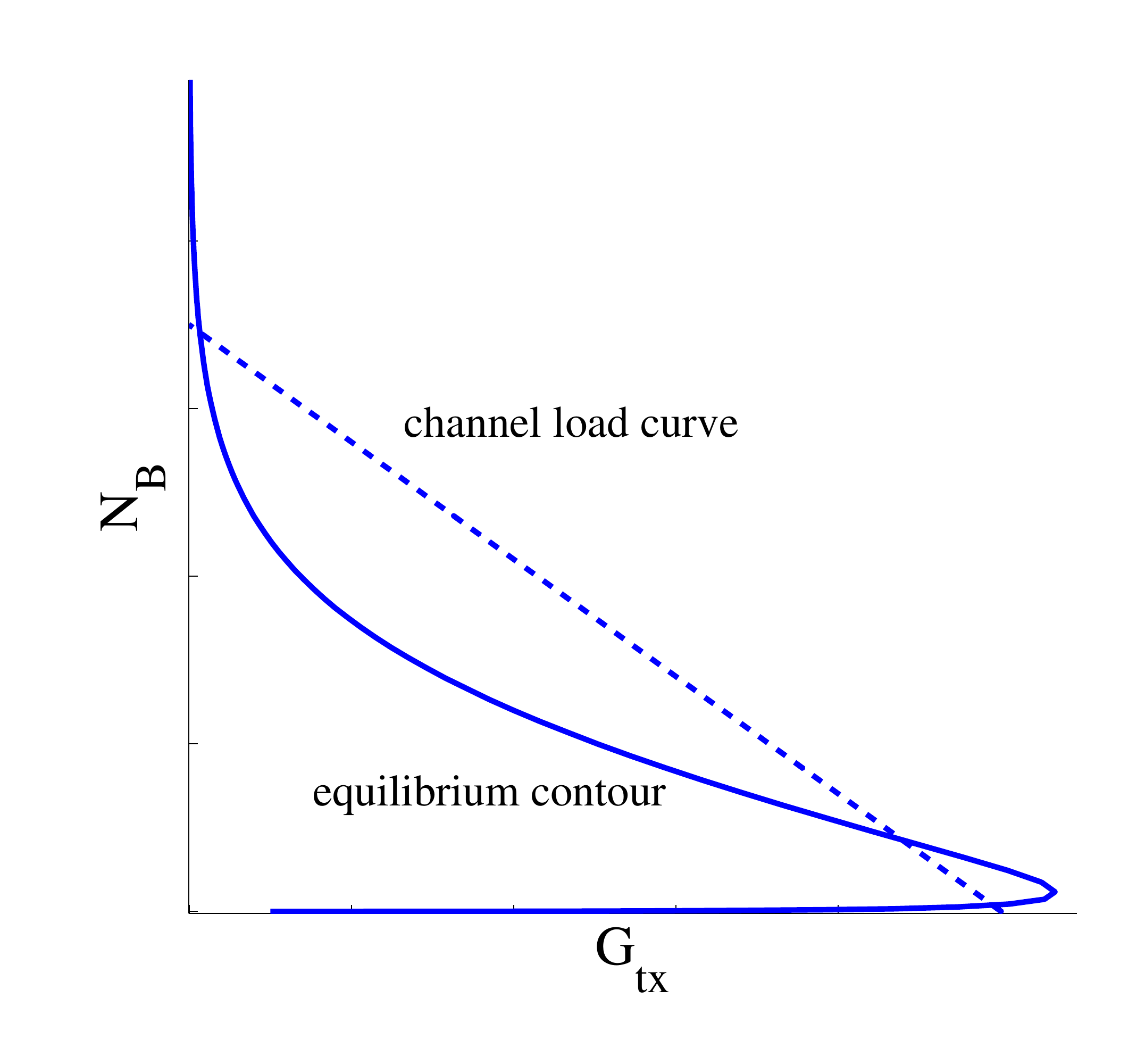}
\caption{Example of equilibrium contour and channel load curve}
\label{EC}
\end{figure}

As an example consider Figure~\ref{EC} representing two curves. The solid one represents the expected throughput as a function of the number of backlogged users $N_B$ under the constraint $G_{tx}(N_B)=T(N_B)$. The dashed curve represents the actual load due to new transmissions depending on $N_B$. The points of intersection represent points of equilibrium and according to the definition above, the two intersections close to the axis are of stable equilibrium while the remaining one is unstable.
Depending on the number of users and on the policies $p^{tx}$ and $p^{retx}$, it is possible to design a communication channel that has a single globally stable equilibrium point for the maximum achievable throughput.
Under these considerations it results clear that even though SW-CRDSA overcomes SA in a narrow interval of channel load values, this can be sufficient to justify its use in a stable channel having its globally stable equilibrium point into that interval.

\section{Conclusions}

In this paper an analysis in terms of normalized efficiency for the recently introduced Sliding Window - CRDSA technique has been presented. The need for such an analysis finds its reason in the use of Contention Resolution Diversity Slotted Aloha as Random Access communication technique for transmission in scenarios with limits on the average power (e.g. transponder's energy limitations in satellite communications). For this reason, a comparison that takes into account fairness in the use of available energy (in this case at the relay) is needed.
Found results clearly show that the use of an unframed access to the channel is more convenient than a division of the channel in frames since Sliding Window - CRDSA outperforms Frame Based - CRDSA regardless of the actual burst degree distribution chosen. Moreover while with Frame Based techniques SNR greater than $6\ dB$ is needed in order to get better results than SA in terms of normalized efficiency, with the use of Sliding Window - CRDSA better results around the throughput peak are already found for $SNR=0\ dB$.
The obtained results find application in open loop scenarios as well as in the case of retransmission of unresolvable packets, under the assumption that the channel is globally stable.\\


\begin{thebibliography}{1}

\bibitem{SWCRDSA:RobertsALOHA}
L.G.~Roberts, "ALOHA packet systems with and without slots and capture", ARPANET System Note 8 (NIC11290), June 1972.

\bibitem{SWCRDSA:AbramsonALOHA}
N.~Abramson, "The throughput of packet broadcasting channels", \emph{IEEE Trans.Comm.}, vol.25, pp.117-128, Jan. 1977.

\bibitem{SWCRDSA:DiversityALOHA}
G.L.~Choudhury and S.~S.~Rappaport, "Diversity ALOHA - A random access scheme for satellite communications", \emph{IEEE Trans.Comm.}, vol.31, pp.450-457, Mar. 1983.

\bibitem{SWCRDSA:CRDSA1}
Casini, E.; De Gaudenzi, R.; Herrero, Od.R.; , "Contention Resolution Diversity Slotted ALOHA (CRDSA): An Enhanced Random Access Schemefor Satellite Access Packet Networks," Wireless Communications, IEEE Transactions on , vol.6, no.4, pp.1408-1419, April 2007

\bibitem{SWCRDSA:CRDSA2}
O. del Rio Herrero and R. De Gaudenzi, "A High-Performance MAC Protocol For Consumer Broadband Satellite Systems", Proceedings of the 27th International Communications Satellite Systems Conference (ICSSC), June 1st-4th, 2009, Edinburgh, Scotland. 

\bibitem{SWCRDSA:CRDSA3}
De Gaudenzi, R.; del Rio Herrero, O.; , "Advances in Random Access protocols for satellite networks," Satellite and Space Communications, 2009. IWSSC 2009. International Workshop on , vol., no., pp.331-336, 9-11 Sept. 2009

\bibitem{SWCRDSA:IRSA1}
Liva, G.; , "Graph-Based Analysis and Optimization of Contention Resolution Diversity Slotted ALOHA," Communications, IEEE Transactions on , vol.59, no.2, pp.477-487, February 2011

\bibitem{SW}
Meloni, A. and Murroni M. ; Kissling, C. and Berioli, M.; , ”Sliding Window-Based Contention Resolution Diversity Slotted ALOHA”, Proceedings of the Global Communications Conference, GLOBECOM 2011, 3-7 December 2012, Anaheim, California, USA.

\bibitem{CRDSA_stab1}
Meloni, A.; Murroni, M.; , "CRDSA, CRDSA++ and IRSA: Stability and performance evaluation", Advanced Satellite Multimedia Systems Conference (ASMS) and 12th Signal Processing for Space Communications Workshop (SPSC), 2012 6th , vol., no., pp.220-225, 5-7 Sept. 2012

\end{thebibliography}
\end{document}